\documentclass[11pt]{article}
\usepackage[font=small,labelfont=bf]{caption}
\usepackage{listings}
\usepackage{subcaption}
\usepackage{amsmath,amssymb}
\usepackage[left=1in, right=1in, top=1in, bottom=1in]{geometry}
\usepackage{verbatim}
\usepackage{titling}
\usepackage{parskip}

\usepackage{todonotes} 

\usepackage[british]{babel}
\usepackage[utf8]{inputenc}
\usepackage{slashed}
\usepackage{amsfonts}
\usepackage{amsthm}
\usepackage{amsmath,amssymb}
\usepackage{empheq}
\usepackage{xcolor}
\usepackage{hyperref}
\usepackage{nag}

\newcommand{\pd}[2]{\frac{\partial#1}{\partial#2}}

\newcommand{\ER}{Erd\H{o}s-R\'enyi }

\usepackage{xcolor}

\usepackage{soul}

\title{PDE-limits of stochastic SIS epidemics on networks}
\author{F. Di Lauro$^1$, J.-C. Croix$^1$,  L. Berthouze$^2$, I.Z. Kiss$^1$}
\date{$^1$ Department of Mathematics, University of Sussex, Falmer, BN1 9QH, UK\\
	  $^2 $ Department of Informatics, University of Sussex, Falmer BN1 9QH, UK\\ 
	  \vspace{1cm}\today
	  }

\begin{document}
\maketitle
\begin{abstract}

Stochastic epidemic models on networks are inherently high-dimensional and the resulting exact models
are intractable numerically even for modest network sizes. Mean-field models provide an alternative but can only capture 
average quantities, thus offering little or no information about variability in the outcome of the exact process.
In this paper we conjecture and numerically prove that it is possible to construct PDE-limits of the exact stochastic SIS epidemics on 
regular and Erd\H{o}s-R\'enyi networks.
To do this we first approximate the exact stochastic process at population level by a Birth-and-Death process (BD) (with a state space of $O(N)$ rather than $O(2^N)$) whose
coefficients are determined numerically from Gillespie simulations of the exact epidemic on explicit networks.
We numerically demonstrate that the coefficients of the resulting BD process are density-dependent, a crucial condition for the existence of a PDE limit.
Extensive numerical tests for Regular and Erd\H{o}s-R\'enyi networks show excellent agreement between
the outcome of simulations and the numerical solution of the Fokker-Planck equations. 
Apart from a significant reduction in dimensionality, the PDE also provides the means to derive the epidemic outbreak threshold linking network and disease dynamics parameters, albeit in an implicit way. Perhaps more importantly, it enables the formulation and numerical evaluation of likelihoods for epidemic and network inference as illustrated in a fully worked out example.
\end{abstract}


\section{Introduction}
An epidemic is a complex phenomenon that arises from a pathogen spreading over the contact structure of a population. Similar spreading phenomena occur in various disciplines, from biology and social sciences to engineering. Unsurprisingly, much modelling effort has been put into studying spreading processes on networks, as they offer a natural framework to mimic real-life contact patterns~\cite{brauer2012mathematical} and the important heterogeneities within these. The use of networks is extremely intuitive with each individual encoded as a node, and all its contacts (to other individuals) as links.
Unfortunately, the resulting exact probabilistic model does not scale well with the size of the network, $N$. Even when relatively simple models, such as susceptible-infected (SI) or susceptible-infected-susceptible (SIS), are considered, the exact model has $2^N$ equations, which quickly becomes intractable. 

To address this high dimensionality, mathematical descriptions often focus on population-level statistics (e.g. expected number of infected people at any given time). 
This has led to a number of so-called mean-field models~\cite{Kiss2017,porter2016dynamical,pastor2014epidemic}, offering a good first approximation of the evolution of some population-level or averaged quantities.
These include pairwise models based on moment closure techniques~\cite{Keeling1999,Kiss2017}, effective-degree~\cite{lindquist2011effective}, edge-based-compartmental~\cite{miller2012edge} models and even PDE models~\cite{silk2014exploring}. All such mean-field models share a number of caveats~\cite{Roberts2015}. For example, (i) in general the agreement between these and the exact stochastic model breaks down close to the epidemic threshold, (ii) there are very few cases where it is possible to prove mathematically that the mean-field model is the limit of the exact stochastic process (this has only been done for SIR epidemics and configuration networks~\cite{decreusefond2012large,janson2014law}) and (iii) they give no estimate of the variability observed in the exact process. It is also well-known that such mean-field models only work for a limited class of networks; epidemics on clustered networks are not well-understood, except for idealised clustered networks.

There are ongoing efforts to try to understand and answer rigorous mathematical questions when it comes to analyse  or approximate dynamical processes on networks, see~\cite{scalinglimitsreport} for a recent summary. Progress in this area is usually achieved by bringing in and combining results and techniques from different areas of mathematics. One particularly promising prospect for SIS epidemics on networks is to consider them as Birth-and-Death (BD) processes. In a recent paper~\cite{Dilauro2019}, we conjectured and confirmed numerically that SIS epidemics are well captured by BD processes, whose rates encode characteristics of both the network structure and the epidemic dynamics. This was tested on Regular, Erd\H{o}s-R\'enyi and Barab\'asi-Albert networks. This choice was motivated by the intuition that epidemic spread is driven by the `birth' of new infected nodes. However, this occurs at a rate which is proportional to the number of S-I (active) links, and these are readily observable during explicit stochastic simulations of the epidemic on networks.

\begin{figure}
\centering
\includegraphics[scale=0.3]{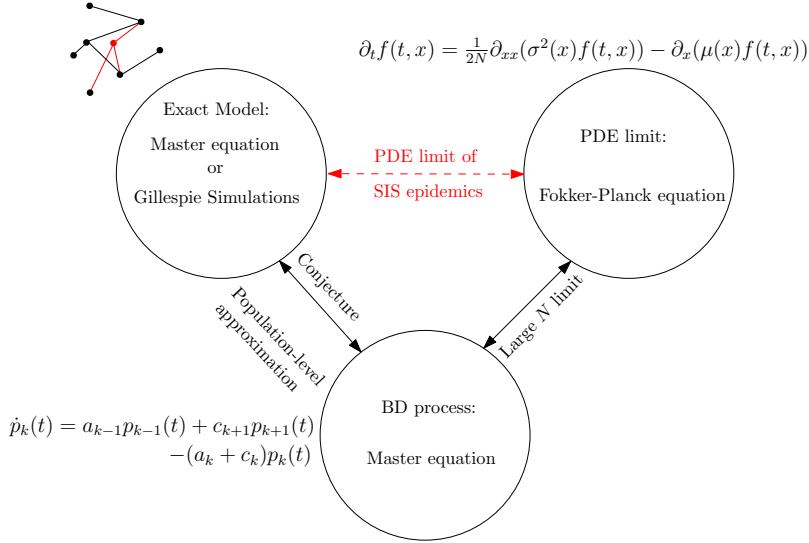}
\caption{Schematic illustration of various approximations of the exact stochastic SIS epidemics on networks. The PDE limit comes as a result, and further confirms the validity, of the Birth-Death approximation conjectured in \cite{Dilauro2019}.}
\label{fig:circle}
\end{figure}

In this paper we build on the above observation and take the next natural step, that is, to consider the large $N$ limit of the BD process, i.e., the  one-dimensional PDE (Fokker-Planck equation). We focus on Regular and Erd\H{o}s-R\'enyi networks and show that the resulting rates in the BD process are density-dependent such that the limit is well defined in the sense of~\cite{vankampenbook}. We compute the rates numerically and also provide a parametric form for them. We show that the resulting PDE agrees well with the output of explicit simulations of stochastic epidemics on networks. The existence of the PDE limit has multiple advantages. First, it reduces further the dimensionality of the system. Second, it gives us the opportunity to compute an epidemic threshold even in an implicit form. Third, it provides the means to get a handle on the variability of the stochastic process with the solution of the PDE providing a likelihood that can computed cheaply and efficiently for inference purposes. Finally, the good agreement between the PDE and the exact process provides further evidence that the BD model may indeed serve as a valid approximation of the exact process (the relation between the exact, BD and PDE-limit model is illustrated in figure~\ref{fig:circle}) and, where a formal proof, beyond numerical validation, may be possible.

This paper is structured as follows. In Section~\ref{sec:Methods} we briefly outline the Birth-and-Death approximation of SIS epidemics as in~\cite{Dilauro2019}.
In Section~\ref{sec:Results} we numerically test and prove that the conditions for the existence of the large $N$ limit, that is, the existence of the PDE, are met. This is done for Regular, Erd\H{o}s-R\'enyi networks. We then show that the solutions of the partial differential equations agree well with the empirical distributions based on simulations of the true process. In Section~\ref{sec:Conclusions} we draw some conclusions and outline further research directions.   
\section{Methods}\label{sec:Methods}
\subsection{Birth-and-Death Approximation of SIS Epidemics}

We briefly describe the model proposed in~\cite{Dilauro2019} and which conjectures that exact stochastic SIS epidemics on networks can be approximated by BD processes. A standard SIS model on an undirected unweighted network $\mathcal{G}$ with $N$ is considered, where each node is either susceptible (S) or infected (I). Infected nodes spread infection to their neighbours with constant per-link rate $\tau$ and recover with rate $\gamma$ (independent of the network). This stochastic process results in a continuous time
Markov chain on a space state of cardinality $2^N$, which forbids analysis even for relatively small values of $N$. Instead, we consider the population-level count of infected nodes, defined as $k(t) = \sum_{i=1}^N \mathcal{I}_{i}(t)$, where $\mathcal{I}_i$ is an indicator function equal to $1$ if node $i$ is infected at time $t$ and $0$ otherwise. $k(t) \in \left[0,N\right]$, where $k(t) = 0$ indicates the state where no infection is present in the network. Given the stochasticity of the process, $k(t)$ is itself a random variable taking values on state space of cardinality $(N+1)$, which is computationally a lot more computationally. We note that each time an infection/recovery occurs, the value of $k(t)$ changes by discrete jumps of size $\pm 1$, respectively. This has led to the conjecture~\cite{Dilauro2019}  that the population-level count process can be approximated by a Birth-and-Death process, governed by the following master equation:
\begin{equation}
\dot{p}_k (t)= a_{k-1}p_{k-1} (t)+ c_{k+1}p_{k+1} (t) - (a_k + c_k)p_k (t),
\label{eq:BDmastereq}
\end{equation}
where $p_k(t)$ is the probability of having $k$ infected nodes at time $t$, $c_k = \gamma k$ is the global recovery rate when $k$ nodes are infected, and $a_k$ is the rate at which the population goes from $k$ to $k+1$ infected individuals. 

The approximation is exact in the case of a complete or fully connected network, where the $a_k$ rates are given by the expression $a_k = \tau k (N-k)$. In the general case, the $a_k$'s are random variables themselves, since the rates at which infections happen are the product of $\tau$ times the total number of $S-I$ links in the network, a random variable that reflects the topology of the network and the way in which the epidemic positions the $k$ infected nodes on the network. This means that the epidemic at population-level is not Markovian, making an exact treatment much more difficult and still out of reach.

However, by using the master equation, we can recast this process as a Markovian one using a suitable approximation of each rate $a_k$. A natural proposal is a quantity that captures the average rate of infection, weighted by the time spent in the observed states, that is:
\begin{equation}
a_k = \frac{\tau \sum_{\xi_k} \xi_k t_{\xi_k}}{\sum_{\xi_k} t_{\xi_k}},
\label{eq:akrates}
\end{equation}
where $\xi_k$ are the observed counts of the number of S-I links on the network when $k$ infected nodes are present and and $t_{\xi_k}$ is the lifetime of this particular state. This quantity 
is responsible for driving the epidemic: The higher the number of S-I links, the larger the rate of generating more infected nodes. The $a_k$'s can be obtained by averaging over many realisations of the epidemic on different realisations of networks from the same family. This can also be interpreted as averaging out stochasticity at link-level and keeping it at population-level. Hence, the variability in epidemic paths will be due to the stochasticity of the master equation itself, guaranteeing the Markov property of the Birth-Death process. The solution of equation~\ref{eq:BDmastereq} with these proposed rates has been shown to be in excellent agreement with the average from many simulations for various network models and epidemics~\cite{Dilauro2019}. \\

\subsection{Fokker Planck equation as a limit of the Birth-Death process}

Master equation~\ref{eq:BDmastereq} can be used as a starting point to build its continuous (in space) limit, i.e., the Fokker-Planck equation~\cite{gardiner2004, Kiss2017}. The idea is to approximate the solution $p_k(t)$ by considering it as a discretisation of a continuous function $f(t,x)$ in the interval $[0,1]$, defined as

\[
f\bigg(t, x = \frac{k}{N}\bigg) = p_k(t).
\]
For the large $N$ limit to exist, it is known~\cite{Kiss2017,kurtz1970solutions,ethier2009markov,Nagy2014,Andras13} that the rates of the master equation need to satisfy the following density condition: 
\begin{equation}
a\bigg(\frac{k}{N}\bigg) = \frac{a(k)}{N}, \; c\left(\frac{k}{N}\right) = \frac{c(k)}{N}.
\label{eq:densitydependence}
\end{equation}
It is worth noting that condition~\eqref{eq:densitydependence} is not guaranteed to hold for any network model, and must therefore be validated on the network models of interest. We expect this condition to hold for networks we refer to as frequency-dependent networks, whereby local topology is preserved as $N$ increases (regular networks, for example). In Section~\ref{sec:Results}, we explore in more detail which network models satisfy this condition.

In the density-dependent case, it can be shown~\cite{Kiss2017,Nagy2014,Andras13} that $f(t,x)$ satisfies the following forward Fokker-Planck equation:
\begin{equation}
\partial_t f(t,x) = \frac{1}{2 N} \partial_{xx}\big(\sigma^2(x)f(t,x)\big) - \partial_x\big(\mu(x) f(t,x)\big),
\label{eq:fokkerplanck}
\end{equation}
with initial condition $f(0,x) = \delta (x-x_0)$, where the diffusion coefficient $\sigma^2(x)$ and the drift $\mu(x)$ are related to the $a_k$ and $c_k$ rates via:
\begin{eqnarray}
\sigma^2(x) &=& \frac{1}{N} (a(x) + c(x)),  \nonumber \\
\mu(x) &=& a(x) - c(x).
\end{eqnarray}
Boundary conditions are naturally emerging from two considerations: (1) if the process hits $k = 0$ at some time (disease-free state) it will stay there forever, (2) the number of infected nodes cannot be greater than $N$ at any given time. Such physical constraints translate naturally in this framework into Dirichlet and Robin boundary conditions:
\[
\begin{cases}
f(t,x=0) = 0, &\text{absorption in $x=0$}, \\
\frac{1}{2} \partial_{x}(\sigma^2(x)f(t,x)) \vert_{x=1} -(\mu(1) f(t,1) = 0, & \text{reflection in } x=1.
\end{cases}
\]
Fokker-Planck equations of this kind have been extensively studied numerically, especially in the biological context of population random genetic drifts~\cite{Duan18,Chen2016,Applegate13,Cacio2012}, as well as analytically~\cite{Feller1954,Trabelsi17,Kovacevic2018}. In particular, in~\cite{Kovacevic2018}, this equation is studied in the limit of large $t$ to characterise the so-called quasi-steady state~\cite{Meleard2012,Collet2013} (where the only steady state possible is absorption at $0$), whereas in~\cite{Cacio2012, Chen2016} various numerical schemes to solve such equations are employed and compared in terms of numerical instabilities and performance. In the Appendix we describe our numerical scheme of choice, which is an adaptation of a finite volume method (FVM) scheme already described in~\cite{Chen2016}.
 \begin{table}[h!]
 	\centering
 	\begin{tabular}{|l|l|l|l|l|}
 		\hline
 		\multicolumn{1}{|l|}{Networks}& $\langle k \rangle$ & $\tau$ & $\gamma$ & $R_0$ \\ \hline
 		&            $9$          &   $1$     &     $6$     &   $1.28$    \\ 
 		\multicolumn{1}{|l|}{Regular} &            $7$          &   $2.5$   &     $8$     &   $1.66$     \\
 		&            $8$          &   $3.5$   &     $7$     &   $2.65$    \\
 		\hline
 		&            $8$          &   $1$     &     $5$     &   $1.33$    \\
 		\multicolumn{1}{|l|}{Erd\H{o}s-R\'enyi}&            $10$         &   $1$     &     $4.5$   &   $1.80$    \\ 
 		&           $7$          &    $4$    &     $7$     &   $2.54$    \\\hline
 	\end{tabular}
 	\caption{Values of network and epidemic parameters for the benchmark scenarios chosen to test the PDE limit of large networks. $R_0$ has been computed using the formula $R_0 = \frac{\tau \langle k \rangle}{\tau + \gamma}$ as described in~\cite{Kiss2017}.}\label{table:eqn}
 \end{table}
 \begin{figure}[h!]
 	\vspace*{-5pt}
 	\centering
 	\includegraphics[scale=1]{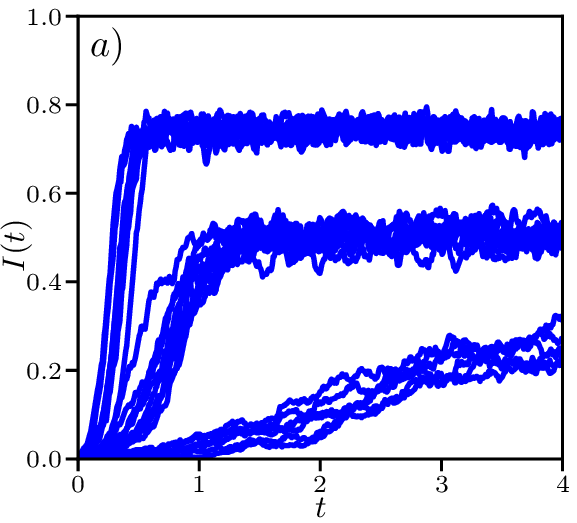}
 	\includegraphics[scale=1]{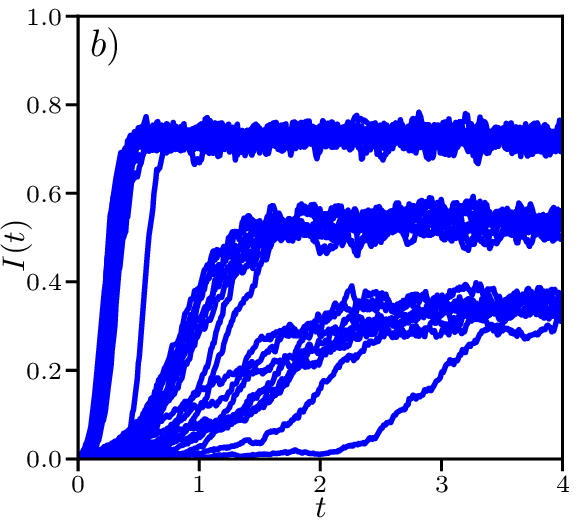}
 	\caption{Typical realisations of SIS epidemics on  (a) Regular and (b) Erd\H{o}s-R\'enyi networks, for the parameter values shown in Table~\ref{table:eqn} and with $N=1000$. In each panel $10$ realisations of the epidemics are plotted,. The parameters used to generate such networks are reported in table~[\ref{table:eqn}], higher epidemics corresponding to higher values of $R_0$.}\label{fig:epiexamples}  
 	\vspace*{-5pt}
 \end{figure}
\section{Results}\label{sec:Results}
\subsection{Validation of the density dependence condition}
\begin{figure}[h!]
	\vspace*{-5pt}
	\centering
	\includegraphics[scale=0.9]{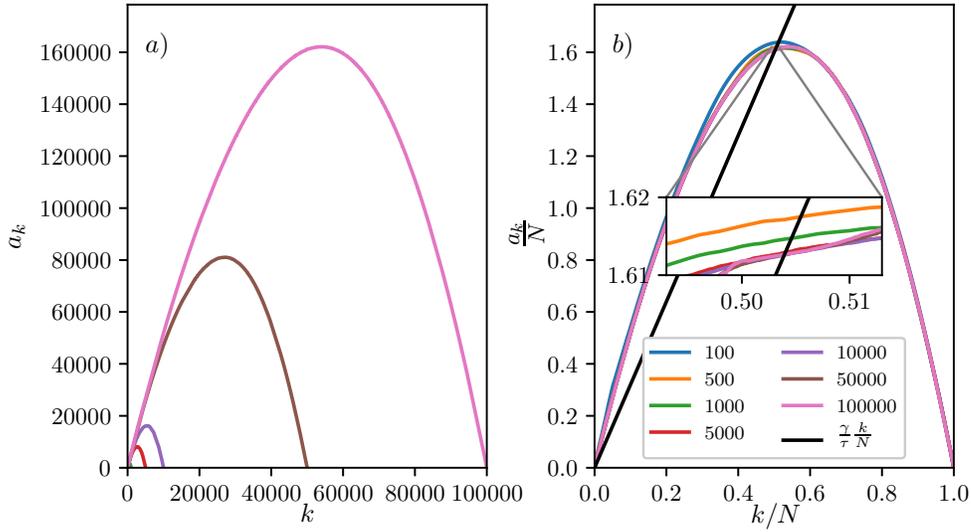}
	\caption{ Scaling for regular networks using parameters given in the first row of Table~(\ref{table:eqn}). (a) Unscaled $(k,a_k)$ curves for values of $N$ ranging from $N=100$ to $N=100000$. Each curve is obtained by simulating $10000$ realisations of the epidemic across $50$ realisations of the network, half of the epidemics starting from $k_0=1$, the other half from $k_0=N$. (b) Corresponding scaled rate $(k,\frac{a_k}{N})$ curves. The scaling hypothesis can be checked by noticing that the higher the values of $N$, the closer the scaled curves get to the limiting universal curve. As $N$ increases, the differences between scaled rates decrease. In the inset, the small mismatch between curves with $N\geq 1000$ are highlighted using a $30$x zoom. For completeness, the ($k/N, \gamma k/N$) curve is also reported (in black); it intercepts the scaled curves around the steady state.}
	\label{fig:regscaling}
	\vspace*{-5pt}
\end{figure}

\begin{figure}[h!]
	\vspace*{-5pt}
	\centering
	\includegraphics[scale=0.9]{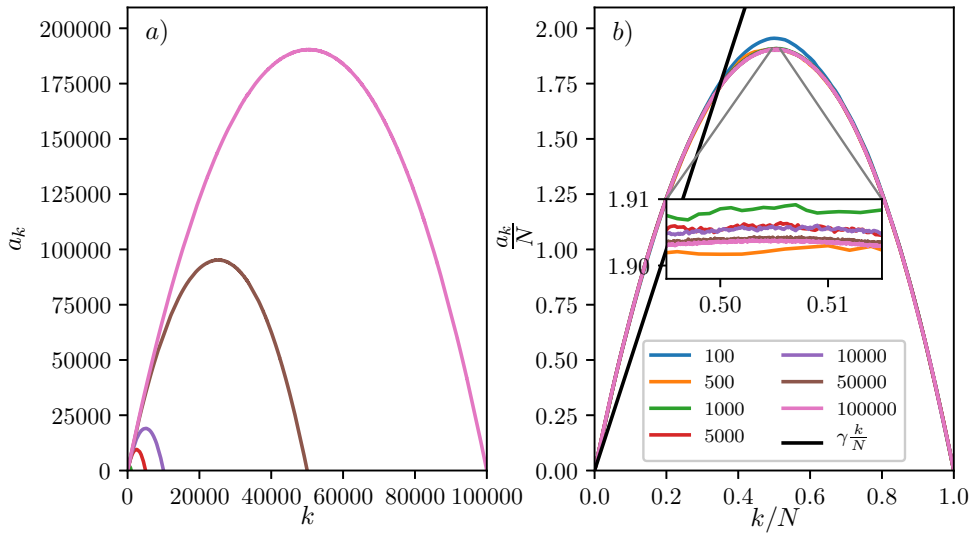}
	\caption{Same scenario as in figure~\ref{fig:regscaling} but for Erd\H{o}s-R\'enyi networks using parameter values from the fourth row of Table~(\ref{table:eqn}). (a) Unscaled $(k,a_k)$ curves for values of $N$ ranging from $N=100$ to $N=100000$. Each curve is obtained by simulating $10000$ realisations of the epidemic across $50$ realisations of the network; half of the epidemics starting from $k_0=1$, the other half from $k_0=N$. (b) Corresponding scaled $(k,\frac{a_k}{N})$ curves.}\label{fig:erscaling}
	\vspace*{-5pt}
\end{figure}

In order to use eq.~\eqref{eq:fokkerplanck} we need to verify that the rates of the BD process satisfy condition~\eqref{eq:densitydependence}. Recoveries are independent of the network, therefore, the condition is automatically satisfied as the expression for these rates is $c_k = \gamma k$. Infection rates, instead, need to be inspected more closely, as their values are dependent on the topology of the network. Indeed, condition~\eqref{eq:densitydependence} has an important physical interpretation: whilst globally epidemics might look very different at different network sizes, locally the network structure may not change. For example if we consider a regular network with degree $m$, each node maintains the same number of neighbours, independent of network size. Thus regular networks are excellent candidates to test the density-dependence assumption. This is similarly true for Er\H{o}s-R\'enyi networks with fixed average degree. 
 On the contrary, if we consider a Barab\'asi-Albert network model, we observe that whilst the average degree remains constant, the hubs become more connected as network size increases, which in turn means that the variance of the degree distribution also increases. This growing influence of the hubs with network size will drastically affect the spreading pattern. For this reason, we do not expect the density-dependent condition to hold for Barab\'asi-Albert networks.
 Finally, fully-connected networks enter this framework by noting that the infection rates have an analytical expression $a_k = \tau k(N-k)$. However, these clearly violate condition~\eqref{eq:densitydependence}. This can be corrected by requiring that $\tau$ scales as $\tau/N=ct$ in the limit of large $N$. This seems to be in line with the frequency-dependent consideration since the rescaling means that the average number of neighbours a node can infect does not simply increase with $N$. This case is well-known in the literature~\cite{Gray2011, Linda17}, albeit in a SDE formulation, so we limit our treatment of it to reporting the exact Fokker-Planck equation for the fully connected network, i.e.
\begin{eqnarray*}
\partial_t f(t,x) = \frac{1}{2 N} \partial_{xx}\left[ \left(\beta x(1-x)  + \gamma x\right)f(t,x)\right] - \partial_x\bigg[ \left(\beta x(1-x) - \gamma x\right) f(t,x)\bigg],
\end{eqnarray*}
 where $\beta = \frac{\tau}{N^2}$.\\
 To test the scaling hypothesis, the infection rates, based on eq.~\eqref{eq:akrates}, computed for different values of $N$ and different networks are plotted in figures~\ref{fig:regscaling} and~\ref{fig:erscaling}. Using eq.~\eqref{eq:densitydependence}, these rates are rescaled and plotted again in the same figures confirming that they define a universal rate.
 When $N$ is scaled, a little variability between the $(k,a_k)$ scaled curves emerges, due to the high degree of stochasticity of the process. However, the difference is so small that the Fokker-Planck equation and its solution appear insensitive to the exact choice of the rates. For here onwards, we use the rates corresponding to the $N=1000$ case. 


\subsection{Comparing PDE and simulations}
Since the limit of large $N$ is of interest it is beneficial to have a continuous function that fits the discrete $a_k$ rates~\eqref{eq:akrates}. In~\cite{Dilauro2019}, the following three-parameter model was proposed:
\begin{equation}
a^{(C,a,p)}_k = C k^p{\big(N-k\big)}^p \bigg(a \bigg(k-\frac{N}{2}\bigg) +N\bigg).
\label{eq:cap}
\end{equation}
This model can be fitted to the $a_k$ curves via a least-square approach, by minimizing the following cost:

\begin{equation}
e = \sum_{k, \sum t_{\xi_k}>0} {( a^{(C,a,p)} - \hat{a}_k )}^2.
\label{eq:ls}
\end{equation}
In~\cite{Dilauro2019}, we showed that this approach leads to good agreement with simulations from different network classes, in particular, regular and \ER networks as considered in this paper.

In the following, we make use of this function to model the infection rates of master equation~\eqref{eq:BDmastereq}.
However, using a simple function to model the complexity of the $a_k$ rates is adding an additional layer of approximation to our approach. Therefore, in addition to eq.~\eqref{eq:cap} we also consider a cubic spline of the $a_k$ rates, as it provides a much better fit to the rates based on eq.~\eqref{eq:akrates} and therefore yields better results. To summarise, the rates of infection are first found based on simulations via eq.~\eqref{eq:akrates}. As this approach produces a discrete function that cannot be used as is in the Fokker-Planck equation, we propose two alternatives: (a) the ($C, a, p$) model, eq.~\eqref{eq:cap}, and (b) a spline. The PDE is considered with both rates and the numerical solution of the PDE is computed via a Finite Volume Method (several other numerical schemes~\cite{masoumeh2015,Cacio2012, Chen2016} were tested) as it guarantees that the solution remains non-negative and preserves mass, see Appendix.

To show the agreement between the Fokker-Planck equation~\eqref{eq:fokkerplanck} and results from simulations on networks, we selected six combinations of network and epidemic parameters, as described in Table~\ref{table:eqn}. We selected three networks from each family (regular and Erd\H{o}s-R\'enyi) and tuned the parameters such that for each family we could get three epidemics with different characteristics, i.e., transient and quasi-steady state. To show this, in figure~(\ref{fig:epiexamples}) we illustrate a few realisations of epidemics on networks of size $N=1000$ for each scenario. Further, we provide the reproduction number $R_0 = \frac{\tau \langle k \rangle}{\tau + \gamma}$\cite{Kiss2017} for each of this scenario. 

Parameters were chosen so that, for each family, the three quasi-steady states showed a prevalence of circa $0.25, 0.5$ and $0.75$, respectively. To find the $(k,a_k)$ curves via minimisation of~\eqref{eq:ls}, we generated data as follows: for each scenario, we created $50$ realisations of the network, and on each we ran $200$ realisations of the epidemic, half of which started from $k_0 = 1$, the other half from $k_0 = N$. This was done in order to obtain observations over the whole range of possible values of the infected nodes. Indeed, when epidemics start from low $k_0$ values, they only very rarely reach a prevalence much higher than the quasi-steady-state. 

\begin{figure}[h!]
	\vspace*{-5pt}
	\centering
	\includegraphics[scale=0.8]{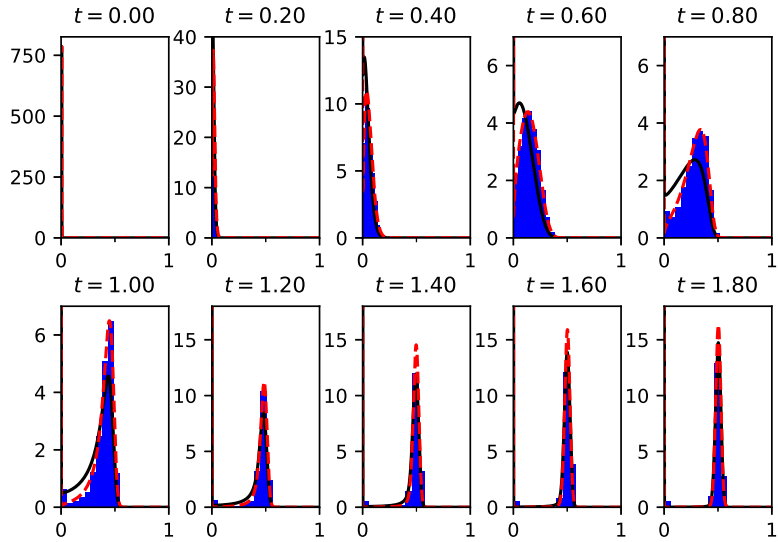}
	\caption{Temporal evolution of the  probability distribution $p_{x=\frac{k}{N}}$(blue histogram) sampled from 25000 realisations of epidemics across 100 realisations of regular networks (2nd row of Table~\ref{table:eqn}), with $N=1000$. Lines are the numerical solutions to the Fokker-Planck equation~(\ref{eq:fokkerplanck}) computed from two different $a_k$ rates: best ($C,a,p$) fit (continuous curve) and cubic spline of the raw $a_k$ computed as in eq.~\eqref{eq:akrates} (dashed line). The first panel shows the initial condition ($t=0$), which for all simulations is $k_0 = 1$, while the last panel shows the quasi-steady state distribution.}\label{fig:PDE_REG}
	\vspace*{-5pt}
\end{figure}

\begin{figure}[h!]
	\centering
	\vspace*{-5pt}
	\includegraphics[scale=0.8]{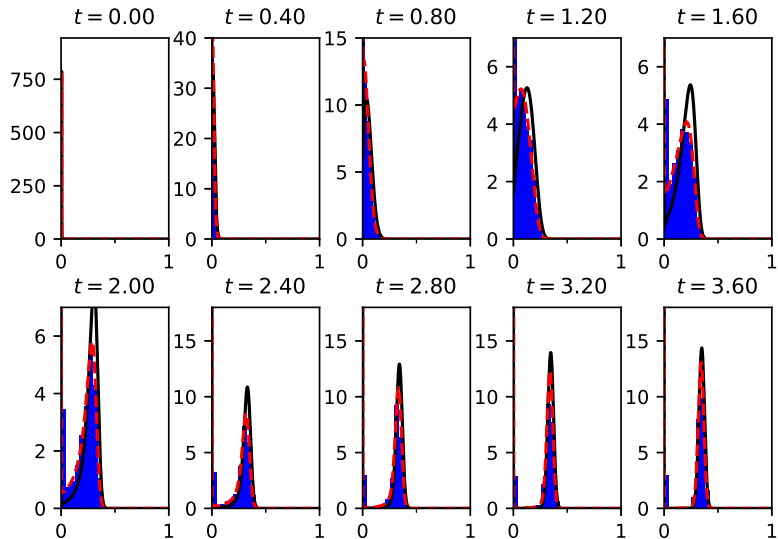}
	\caption{Same scenario as in figure~\ref{fig:PDE_REG}, but using the parameters given in the fourth row of Table~\ref{table:eqn}, i.e., the first parameter configuration for Er\H{o}s-R\'enyi networks.}\label{fig:PDE_ER}
	\vspace*{-5pt}
\end{figure}

The numerical solutions of equation~\eqref{eq:fokkerplanck} are compared with results based on Gillespie simulations~\cite{Gillespie1976,Gillespie1977}, see figures~\ref{fig:PDE_REG} and~\ref{fig:PDE_ER}. The same quality of agreement holds for all  scenarios we tested, as long as the size of the network is $\geq1000$. For small networks, there is a finite-size effect that does not allow for as good a fit. Interestingly enough, although there are small differences between different $a_k$ curves, as long as $N \geq 1000$, the exact choice of $N$ has little impact on the numerical solutions of the PDE. This supports our conjecture that there is indeed a large $N$ limit and, therefore, a universal scaled $a_k$ curve which is approached as $N$ increases. As can be seen, the spline consistently leads to a better approximation. This is simply due to a tighter fit to the discrete data compared to the fit based on the ($C,a,p$) model. We note, however, that the ($C,a,p$) model captures the trend of the epidemic and the quasi-steady state is fitted well.  

To realise the comparisons provided in figures~\ref{fig:PDE_REG} and~\ref{fig:PDE_ER} we proceeded as follows. We considered the same network realisations and epidemic parameters used to find the $(k,a_k)$ rates. We fixed the initial condition to be $k_0 = 1$ and ran $200$ simulations on each realisation. Each individual path was then sampled at regular times in order to build the empirical distribution $p(x,t)$. Note that all simulations were kept, even those that died out early. This is because the numerical scheme preserves the total probability and can account for these early extinctions. 

The PDE with the ($C, a, p$) model is
\begin{eqnarray*}
	\label{eq:PDE_CAP}
	\partial_t f(t,x) = \frac{1}{2 N} \partial_{xx}\bigg[\bigg(CN^{2p}\left(x^p(1-x)^p\right)\left(a \left(x-\frac{1}{2}\right) +1\right) + \gamma x\bigg)f(t,x)\bigg] + \nonumber \\
	- \partial_x\bigg[\bigg(CN^{2p}\left(x^p(1-x)^p\right)\left(a \left(x-\frac{1}{2}\right)+1\right) - \gamma x\bigg) f(t,x)\bigg].
\end{eqnarray*}

Our three-parameter model, ($C,a,p$), can be used to derive the epidemic threshold.
In terms of the PDE, see equation~\ref{eq:PDE_CAP}, and as figures~\ref{fig:PDE_REG} and~\ref{fig:PDE_ER} show, an epidemic is supercritical when the drift term is positive. This implies that the epidemic threshold is equivalent to 
\[
CN^{2p}\bigg(x^p(1-x)^p\bigg)\left(a \left(x-\frac{1}{2}\right)+1\right) - \gamma x \ge 0,
\]
at the start of the epidemic, that is $x \simeq 0$. The fact that the values of $p$ are typically close to one leads to
\[
x\left[CN^{2}(1-x)\left(a \bigg(x-\frac{1}{2}\bigg)+1\right) - \gamma \right] \ge 0,
\]
taking the limit $x\to 0$ in the expression within the brackets above leads to
\[
CN^2\left(1-\frac{a}{2}\right) > \gamma.
\]
This expression reduces to the well-known condition $R_0 = \frac{\tau}{\gamma} \geq 1$ for fully connected networks. Indeed, scaled rates for such networks can be computed exactly to be $a(x) = \frac{\tau}{N^2} x (1-x)$, meaning that $C =\frac{\tau}{N^2}$, $a=0$ and $p=1$ in this case.  

This equation is implicit, as, of course, both $C$ and $a$ depend on the network and epidemic parameters in a non-trivial way. Therefore it cannot be used as it is, but it offers an interesting interpretation since $a$ determines whether the $(k,a_k)$ curves are left- or right-skewed, see~\cite{Dilauro2019}. Furthermore, the topology of the underlying network plays an important role in determining the shape of this curve; for example, Barab\'asi-Albert networks lead to $(k,a_k)$ curves with a left skew~\cite{Dilauro2019}. Thus, all else being constant, networks with high degree heterogeneity are more likely to see the threshold go past the critical value.
\subsection{Inference of infection rates using the Fokker-Planck approximation}
\label{sec:inference}

\begin{figure}
	\centering
	\includegraphics[scale=1]{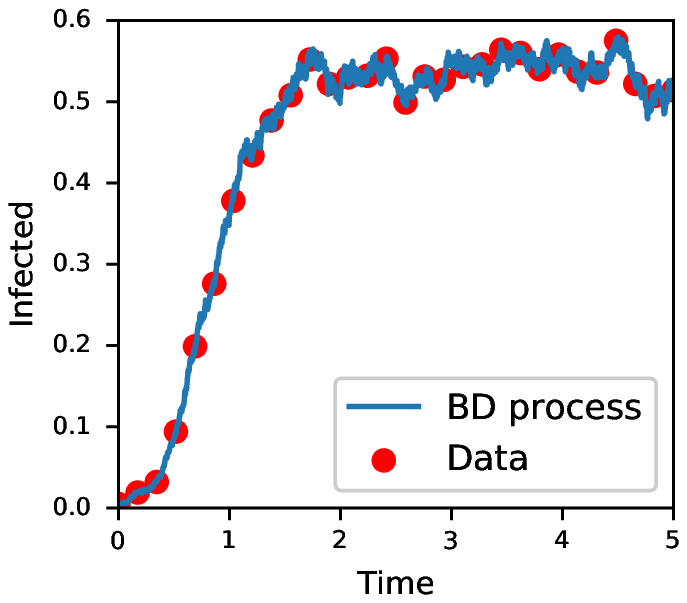}
	\includegraphics[scale=1]{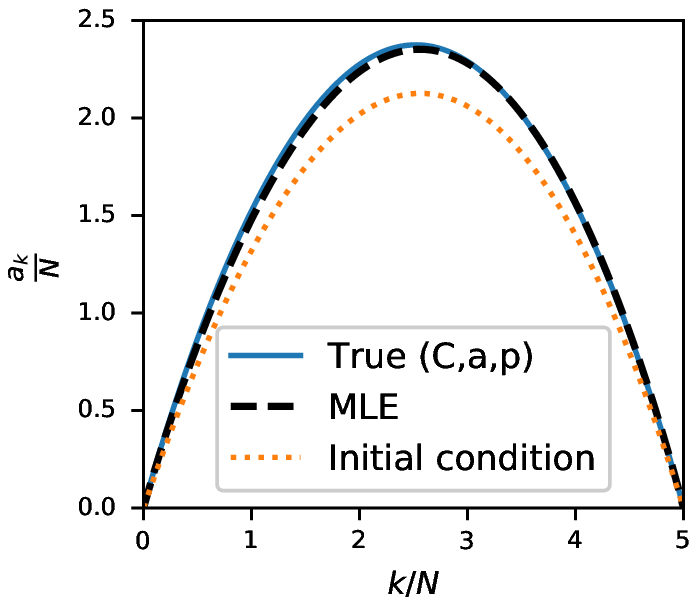}
	\caption{(left) Data generated from a single realisation of an SIS process on an \ER network with $N=1000, \langle k \rangle = 10$, $\tau = 1$, $\gamma=4.5$ via the Gillespie algorithm. (right) $C,a,p$ function obtained by maximising the logarithm of the likelihood~\ref{eq:LikelihoodFP} (black dashed line) compared to the $C,a,p$ curve obtained by fitting the $(k,a_k)$ curve (blue continuous line). The initial condition inputed to the locally bounded gradient-descent solver is shown by the orange dotted line.}
	\label{fig:inference}
\end{figure}

In this last section, we provide a simple example of the usefulness of the Fokker-Planck approximation: inferring epidemic and network parameters given data. Specifically, we consider the case in which a single trajectory of BD (or Gillespie simulation of the epidemic on an explicit network) process is observed at discrete time-steps, i.e.:
\begin{equation*}
	y=\left\lbrace (k_1,t_1),\dots,(k_n,t_n)\right\rbrace,
\end{equation*}
where $(k_1,\dots,k_n)\in\lbrace 0,\dots,N\rbrace^n$ and $(t_1,\dots,t_n)\in[0,T]^n$ are the sets of states and times ($0\leq t_1<\dots<t_n\leq T$), respectively. To set up the inference, we express the likelihood using the transition probability function of a BD process as follows (using the independence of increments and time homogeneity):
\begin{equation*}
	\mathcal{L}_{BD}\left(y;a,c\right)=\prod_{i=1}^{n-1}\mathbb{P}\left(X(t_{i+1}-t_{i})=k_{i+1}\vert X(0)=k_i;a,c\right).
\end{equation*}
Unfortunately, for a large state space, these transition probabilities are numerically expensive to compute. Additionally, inferring the full set of rates $a,c$ may not be efficient. Instead, we recast this problem as that of inferring the $C,a,p$ parameters of the Fokker-Planck approximation. This is a much more tractable numerical problem, that can still provide useful information about the underlying network and epidemic, as showed in~\cite{Dilauro2019}. This amounts to replacing the previous likelihood function with the following:
\begin{equation*}
\mathcal{L}_{FP}\left(y;C,a,p\right)=\prod_{i=1}^{n-1}f(t_{i+1}-t_{i},x_{i+1};x_{i},C,a,p),
\label{eq:LikelihoodFP}
\end{equation*}
where $f(t,x;x',C,a,p)$ is the transition probability density obtained from equation~\eqref{eq:fokkerplanck}, the coefficients are given by the $C,a,p$ model, $x_i=\frac{k_i}{N}$ for all $i\in[1,n]$ and the initial data is a Dirac delta at location $x'\in(0,1)$. In this example, $f$ is computed numerically using the finite-volume numerical scheme described in the Appendix. To illustrate the accuracy of this approach, we consider a set of parameters from the choices of Table~\ref{table:eqn} (figure~\ref{fig:inference} shows the behaviour of the system when parameters are those on the 5th row of Table~\ref{table:eqn}, i.e. $C=1.36e-05,a=3.44e-2,p=9.7e-1$) and generate a trajectory from a single realisation of the SIS epidemic on a Erd\H{o}s-R\'enyi network of size $1000$, via Gillespie algorithm. This dataset is shown in figure~\ref{fig:inference} and consists of $n=30$ distinct data points taken from the epidemic curve. These are then scaled to $[0,1]$ (taking $x_i=\frac{k_i}{N}$ for all $i\in[1,n]$). The dataset is then used to find a Maximum Likelihood Estimator (MLE) by simply maximizing the likelihood function from equation~\eqref{eq:LikelihoodFP} with respect to $C,a,p$, that is finding 
\begin{equation*} 
	(\hat{C},\hat{a},\hat{p})=\arg\max\mathcal{L}_{FP}\left(y;C,a,p\right).
\end{equation*}

To show that this method provides a good estimate, we simply plot the MLE rates, $(\hat{C},\hat{a},\hat{p})$, against the rates that were used to simulate the data in figure~\ref{fig:inference}. The true and estimated rates are indeed= in good agreement. We repeated the inference scheme for all benchmark cases in Table~\ref{table:eqn} (not shown). The agreement was similar to that shown in figure~\ref{fig:inference} (right panel).

  It is worth noting that the goodness of inference depends on how many points the dataset contains and also how much of the transient and the quasi-steady state is captured.  In the transient, the drift dominates and the process is more stochastic. On the other hand, in the quasi-steady state, the drift coefficient tends to $0$ and fluctuations around it are mainly due to diffusion. Hence, both regimes are needed if drift and diffusion are to be inferred correctly. Data in the transient or in the quasi-steady state alone can lead to sub-optimal inference as different parameter combinations that provide good fit can be found.
The example reported in figure~\ref{fig:inference} is an illustration of how useful the PDE limit of epidemics on finite networks can be in a network and epidemic inference setting. Further, the approximation that we provide can also be used in a Bayesian approach, by first setting a prior over the parameters $C,a,p$ for instance. 

\section{Conclusions}\label{sec:Conclusions}

In this paper we conjectured the existence of PDE limits for exact SIS epidemics on regular and \ER networks. The key to our approach is to use a BD approximation which then has a PDE limit provided that the coefficients of the BD process are density-dependent. Hence, one of the main challenges was to verify, at least numerically, that this was the case. We could do this for regular and \ER networks and argued that `frequency-dependent' networks are likely to satisfy this condition without imposing further scaling on other parameters.
Intuitively this means that simply increasing the number of nodes in a network will not change what a node experiences locally, e.g. the number or distribution of neighbours it has. 


%


To solve the PDE numerically, we employed a second order in time finite volume method whose stability was proven in~\cite{Chen2016}. We compared such numerical solutions to probability distribution sampled from the Gillespie simulation. The agreement is in general good and, as expected, it becomes excellent as $N$ increases. The existence of the PDE limit is not surprising, given that the coefficients of the BD process are density dependent. However, it is important to note that the agreement between the solutions of the PDEs and empirical distributions based on simulations provides strong support for the validity of the BD process, strengthening the evidence provided in~\cite{Dilauro2019}, and thus closing the loop illustrated in figure~\ref{fig:circle}.

A PDE perspective on epidemics provides several efficiency gains. The first is to do with computational efficiency and the possibility to quantify variability. More importantly perhaps, the solution of the PDE serves as a likelihood which can be very efficiently computed/evaluated and can form the basis of many networks and epidemic inference models, see Section~\ref{sec:inference}. This is in contrast with approaches where the networks are explicitly modelled~\cite{ma2019inferring} and computational complexity can make inference out of reach.

At least two separate avenues of future research emerge. First, and perhaps most importantly, a theoretical justification for the Birth-and-Death approximation is still needed, if indeed that is possible. Second, there is a need to investigate the extent to which 
this method can be extended to other network families and epidemic dynamics. 
Nevertheless, given that handling exact epidemic models on networks is still challenging even for networks of modest size, we believe that proposing new ways to approximate epidemics is worthwhile and may contribute new modelling and analysis perspectives. 


\section*{Appendix: numerical method for solving the PDE}\label{sec:Numericalmethods}
In this section we detail the numerical method and algorithms used to solve eq.~\eqref{eq:fokkerplanck}. The algorithm employed is an adaptation and modification of the finite volume method (FVM) named FVM3 in~\cite{Chen2016}. 
First, we write the Fokker-Planck equation in the form
\[
\pd{f(x,t)}{t} + \pd{j(x,t)}{x}=0,
\]
where in our case the current term is $j(x,t) = -\frac{1}{2N} \pd{\sigma^2(x)f(x,t)}{x} + \mu(x)f(x,t)$, while  initial and boundary conditions are:
\[
\begin{cases}
f( x,0 ) = \delta ( x-x_0 ), & \text{initial condition},\\
f( 0,t ) = 0, & \text{absorption in $x=0$},\\
\pd{f( x,t )}{x}\vert_{x=1} = 0, & \text{reflection in $x=1$}.\\
\end{cases}
\]
In our case, both $\mu(x)$ and $\sigma(x)$ vanish at $0$, indicating that the only possible steady state is absorption~\cite{Kovacevic2018}. Therefore, the solution to this equation is such that $\lim_{t\to\infty} f(x,t) = \delta(x)$. Further, since the solution should provide a probability density function, we require that $f(x,t) \geq 0$ everywhere (positivity) and that $\int_0^1 f(x,t) \,dx = 1$ for any $t>0$ (conservation of mass).
Finite Volume Methods are a class of numerical methods to solve PDEs~\cite{Eymard2000} in which the constraints  described above are explicitly satisfied, therefore FVM is the natural candidate for this type of problems. Following notation of~\cite{Chen2016} we consider a uniform grid, with spacing $h = \frac{1}{M}$ and grid points $x_i = ih$, $0\leq i \leq M$. Similarly for time, we consider a uniform grid with spacing $\tau$ and grid points $t_i = n \tau$, $0 \leq n \leq n_{max}$.  We define $j_i^n$ and $f_i^n$ to be the numerical approximations of $j(x_i,t_n)$ and $f(x_a,t_n)$, respectively. The control volume $\mathcal{D}_i = \{ x \; s.t.\; x_{i-\frac{1}{2}}\leq x \leq x_{i+\frac{1}{2}}\}$ is associated to each inner point $x_i$, whereas two control domains  $\mathcal{D}_0 = \{ x \; s.t.\; 0 \leq x \leq x_{\frac{1}{2}}\}$ and $\mathcal{D}_M = \{ x \; s.t.\; x_{M-\frac{1}{2}}\leq x \leq 1\}$ are reserved for boundary points. 

First, we define the numerical equations imposed by the boundary conditions. The stability of the numerical scheme (in particular, conservation of mass) is influenced by the boundary condition at $x=0$. Naturally, this condition would be $f(0,t)=0$ (absorption), as we already discussed. However, changing it to be a zero-current condition (i.e. $j(x,t)=0$) results in a numerical solution that is more stable.  This change of condition does not influence the solution, as the discretised process is never evaluated at $x=0$. Therefore, we use the following boundary conditions: 
\begin{equation}
j(0,t) = j(1,t) = 0\label{eq:boundary},
\end{equation}
which translates to:
\begin{equation}
\pd{f(x,t)}{t} + \frac{j_{\frac{1}{2}}}{h/2} = 0, \hspace{2cm} \pd{f(x,t)}{t} - \frac{ j^n_{M-\frac{1}{2}}}{h/2}=0.
\end{equation}
The discretisation of the time derivative can be done with a first-order scheme (as in~\cite{Chen2016}) or a higher-order scheme. We opted for a second-order scheme for time, for which, in general
\[
\pd{f(x_i,t_{n+1})}{t} \approx \frac{3f(x_i,t_{n+1}) - 4f(x_i, t_n)+ f(x_i,t_{n-1})}{2\tau},
\]
and the first iteration is done with the first order time scheme $\pd{f(x_i,t_{n+1})}{t}\approx \frac{f(x_i,t_{n+1})- f(x_i,t_n)}{\tau}$. The reason for this choice is that in our case the current term contains both first and second order space derivatives, so to balance out the required space precision, we matched it with a second-order discrete time derivative. This improved the stability of the solution.

The difference between instances of FVMs is how the current term is discretised. In~\cite{Chen2016}, several different schemes are explored. In particular, the FVM that performed better was the so-called central scheme, in which

\begin{equation}
j^{n}_{i+\frac{1}{2}} = -\frac{1}{2N} \frac{\sigma^2(x_{i+1})f^n_{i+1} - \sigma^2(x_i) f^n_i}{h} + \frac{\mu(x_{i+1})f^n_{i+1}-\mu(x_i)f_i^n}{2},
\end{equation}
where $\sigma^2(x_i) = a(x_i) + \gamma x_i$ and $\mu(x_i) = a(x_i) - \gamma x_i$. 

The initial condition $ f(x,0) = \delta(x-x_0)$ is approximated by a Normal distribution $f(x,0) \approx \mathcal{N} \left(x_0, \tilde{\sigma}\right)$ with $\tilde{\sigma} \ll 1$. The stability of the solution with respect to the variance $\tilde{\sigma}$ is discussed in~\cite{Chen2016}, and we have chosen $\tilde{\sigma} = h^2$. The mismatch that can be seen in figures~\ref{fig:PDE_REG} and ~\ref{fig:PDE_ER} at $0$ is due to the fact that the algorithm cannot reproduce a $\delta$ in $0$, and should not be considered a problem, as the mass outside of $0$ is correctly computed by the numerical solver. To test whether absorption at $0$ could have been a problem for the solver, we repeated the calculation allowing for a small re-infection rate at $0$ $\epsilon >0$, without noticing differences in the results.
Our implementation is available online at https://github.com/Fdl1989/PDElimitofepidemics.

\section{Acknowledgments}
All authors acknowledge support from the Leverhulme Trust for the Research Project Grant RPG-2017-370. The authors acknowledge useful discussions with Dr M. Dashti during the development of this research.
\clearpage
\bibliography{Francesco}
\bibliographystyle{plain}
\end{document}